\documentclass[twocolumn]{aastex63}
\usepackage{graphicx}
\usepackage{float}

\submitjournal{ApJ}
\shorttitle{HST Proper Motions of Andromeda~V and Andromeda~VI}
\shortauthors{Casetti-Dinescu et al.}

\begin{document}

\title{HST Proper Motions of Andromeda~V and Andromeda~VI}

\correspondingauthor{Dana I. Casetti-Dinescu}
\email{casettid1@southernct.edu,dana.casetti@gmail.com}

\author[0000-0001-9737-4954]{Dana I. Casetti-Dinescu}
\affiliation{Department of Physics, Southern Connecticut
  State University, 501 Crescent Street, 
New Haven, CT 06515, USA}
\affiliation{Astronomical Institute of the
  Romanian Academy, Cutitul de Argint 5, Sector 4, 
Bucharest, Romania}
\author[0000-0002-9197-9300]{Marcel S. Pawlowski}
\affiliation{Leibniz-Institut f\"{u}r Astrophysik (AIP),
  And der Sternwarte 16, D-14482 Potsdam, Germany}
\author[0009-0001-3739-7051]{Terrence M. Girard}
\affiliation{Department of Physics, Southern Connecticut
  State University, 501 Crescent Street, New Haven, CT 06515, USA}
\author[0000-0002-2483-2595]{Kosuke Jamie Kanehisa}
\affiliation{Leibniz-Institut f\"{u}r Astrophysik (AIP),
  And der Sternwarte 16, D-14482 Potsdam, Germany}
\affiliation{Institut für Physik und Astronomie, Universit\"{a}t 
  Potsdam, Karl-Liebknecht-Straße 24/25, D-14476 Potsdam, 
  Germany}
\author{Max Martone}
\affiliation{Department of Physics, Southern Connecticut
  State University, 501 Crescent Street, 
  New Haven, CT 06515, USA}
\author{Alexander Petroski}
\affiliation{Department of Physics, Southern Connecticut
  State University, 501 Crescent Street, 
  New Haven, CT 06515, USA}
%\author[0000-0003-0218-386X]{Vera Kozhurina-Platais}
%\affiliation{Space Telescope Science Institute, 3700 San Martin Drive, Baltimore, MD 21218, USA}
%\author[0000-0003-2599-2459]{Imants Platais}
%\affiliation{Department of Physics and Astronomy, Johns Hopkins
%  University, 3400 North Charles Street, 
%  Baltimore, MD 21218, USA}

\begin{abstract}

  We measure the absolute proper motions of Andromeda~V (And\,V)
  and Andromeda~VI/Pegasus (And\,VI) dwarf galaxies, satellites of
  M31 located near its galactic plane. And\,VI is located the farthest
  from M31 among the six satellites with currently
  measured proper motions.

  A combination of ACS/WFC and WFPC2 exposures are utilized, 
  spanning a $20$-year time baseline.
  The WFPC2 exposures are processed using a recently developed deep-learning
  centering procedure as well as the most up-to-date
  astrometric calibration of the camera.
  We use on the order of 100 background galaxies per satellite to
  determine the correction to absolute proper motion.
  For And\,V we obtain an absolute proper motion of
$(\mu_{\alpha} , \mu_{\delta})_{And\,V} = (26.1\pm21.5, -74.2\pm19.1)~\mu$as yr$^{-1}$.
For And\,VI we obtain an absolute proper motion of
$(\mu_{\alpha} , \mu_{\delta})_{And\,VI} = (-1.6\pm12.3, -52.6\pm11.2)~\mu$as yr$^{-1}$.

  Orbit integrations and analyses are made for these two Andromeda satellites using
  two estimates of both the mass and proper motion of M31. It is found that
  And\,V has an orbit consistent within errors
  with alignment with M31's disk and counter orbiting it, although
  this alignment is not well constrained. And\,VI's orbit is better
  determined and is very much consistent with co-orbiting
  with M31's disk. While currently at a distance of $\sim 280$ kpc from M31, And\,VI
  will remain beyond a distance of $\sim 90$ kpc from M31, thus experiencing low
  tidal influence compared to the other M31 satellites with known orbits.
  Both satellites are determined to be well-bound to M31.

\end{abstract}


\section{Introduction \label{sec:intro}}

It has been established that both the Milky Way (MW)
and Andromeda (M31) galaxies
host satellite systems that are not spatially isotropic.
Specifically, they display thin planes of satellites
--- with kinematic coherence --- 
that are tens of degrees away to nearly
perpendicular to the galaxy's own disk.
To firmly establish the kinematic coherence one
needs 3D velocities, and much of this work
has been done for the MW thanks to high-quality,
space-based proper motions
(see e.g., recent review by \citet{Dol2025}
and references therein).
A step further beyond their 3D kinematical coherence
is the issue of the planes' dynamical stability
in time, an issue that has direct bearing on the
formation of such ``disks'' of satellites.
This aspect requires ever more precise proper motions,
but also distances and well-known gravitational
potentials of the host galaxy \citep{Kumar2025}.

Measures of 3D motions of satellites that
{\bf do not} belong to these thin planes are of
value as they contribute to unraveling the
formation of the entire system with its specific
accretion history as well as helping to constrain the
gravitational potential of the host galaxy.
Finally, the orbit history is useful in quantifying the
contribution of tides on the mass profile \citep{Dol2025}
with implications for the star-formation history of the
dwarf galaxy \citep{Savino2025}.

To this end, we continue our program \citep{casetti2024b}
to measure proper motions of M31 satellites
taking advantage of unprecedented
20-year baseline observations.
Here, we present the measurement of the
absolute proper motions
of two M31 satellites, namely And\,V and
And\,VI, and analyze their orbits.
These are the first two M31 satellites
far from the Great Plane of Andromeda (GPoA)
\citep{conn2013,ibata2013,savino2022}
that now have measured proper motions and orbit calculations.
And\,VII is also a satellite with a recently measured proper
motion \citep{warfield2023} based on $\sim 30$ {\it Gaia} stars,
but no orbit calculation was performed for it.
Among M31's satellites with proper-motion measurements,
And\,VI is also the most distant from M31 located
some 280 kpc from its center \citep{savino2022}.
As such, it is interesting to determine its
orbit in the context of its recent
mass profile mapping by \citet{Pickett2025}.
Also notable is the fact that these two satellites
lie close to the disk of M31 (i.e., would be
hidden behind the disk, if viewed from a similar position
within M31 that we have in the MW disk).

With this study, we increase the sample of M31
satellites with measured proper motions
to six,
%\footnote{If we also consider M33 and IC10
%  \citep{Brunthaler2005,Brunthaler2007,vandermarel2019,Bennet2024}
%  then the number of M31 satellites is seven.}
including And\,III \citep{casetti2024b}, And\,VII \citep{warfield2023},
NGC\,147 and NGC\,185 \citep{sohn2020}.
Additionally, if we include M33 and IC10 as part of the M31 system
\citep{Brunthaler2005,Brunthaler2007,vandermarel2019,Bennet2024},
there are eight such systems.
  
We analyze the orbits of And\,V and And\,VI using two
mass models for M31, and two proper-motion determinations
of M31. Updated RR-Lyrae-based distances \citep{savino2022}
are also adopted.

The proper-motion measurements are described in
Sections \ref{sec:proc},
\ref{sec:pms} and \ref{sec:absolute}.
Orbit calculations and analysis
are presented in Section \ref{sec:orb-ana}.
Concluding remarks are given in Section \ref{sec:sum}.

\section{Data Processing \label{sec:proc}}
Our proper-motion study is based on HST observations from two epochs,
namely $\sim 2000$, and $\sim 2020$. For each satellite, the early
epoch consists of 24 WFPC2 exposures while the late one consists of
22 ACS/WFC exposures. Image data were downloaded from the Mikulski Archive for
Space Telescopes (MAST). Characteristics of these exposures
are given in Table \ref{tab:prop-data}, including the original observing
proposal id number (PID) for reference.

\begin{figure*}
    \centering
    \includegraphics[scale=0.45,angle=-90]{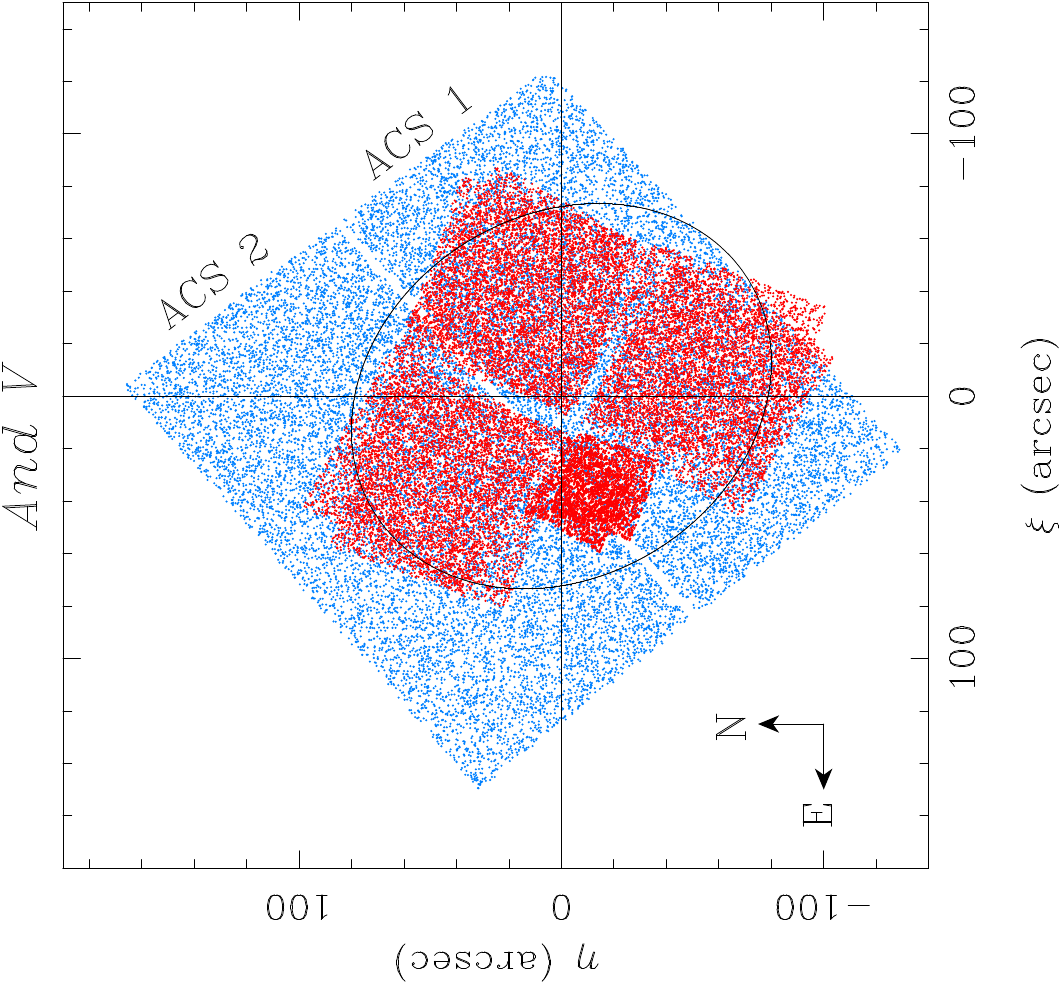} 
    \includegraphics[scale=0.45,angle=-90]{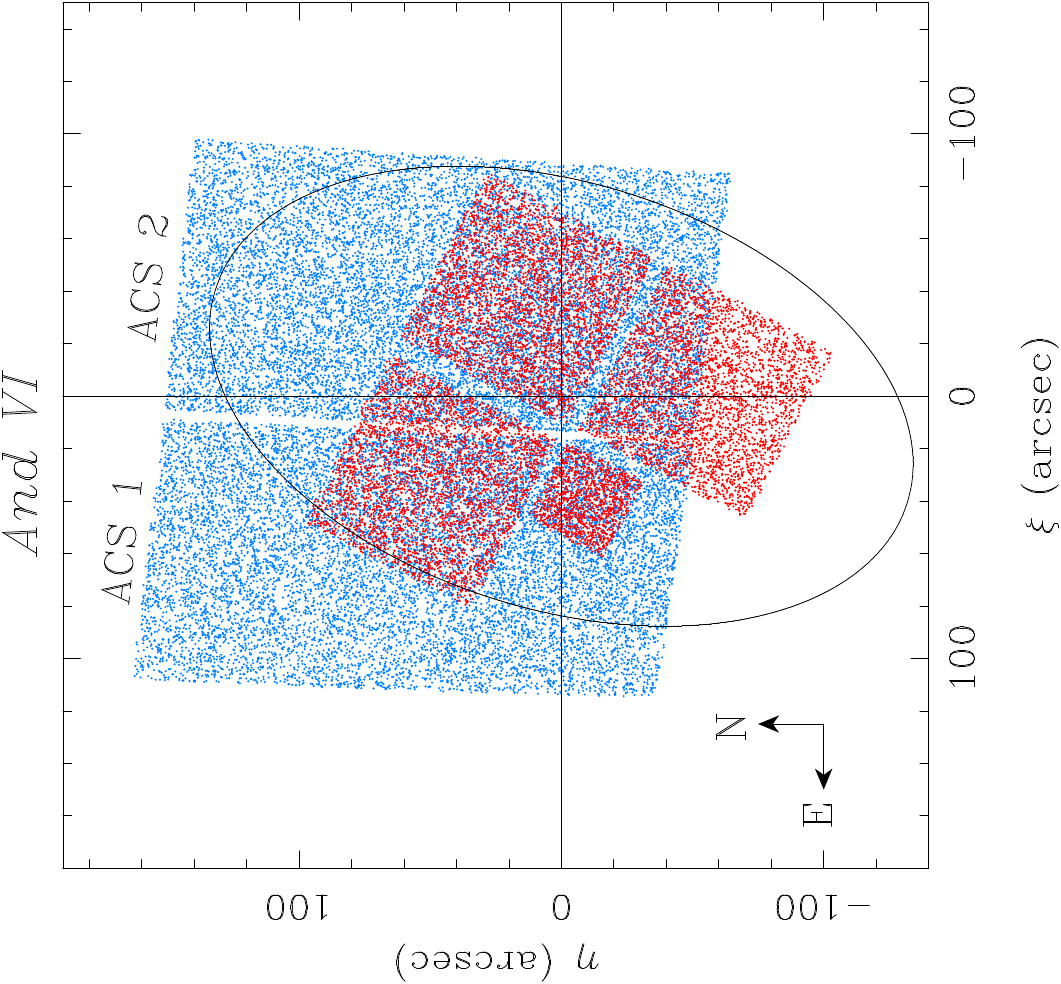}
    \caption{Overlap between the early WFPC2 exposures (red symbols) and
      the late ACS/WFC exposures (blue symbols).
      Coordinates are a gnomonic projection of RA and DEC
      with the tangent point $(\xi, \eta) = (0,0)$ at
      satellite's center. The one half-light ellipses are also
      shown according to the parameters in
      \citep{mcconn2012}.}
    \label{fig:overlap}
\end{figure*}

In Figure \ref{fig:overlap} we show the
overlap between the
WFPC2 exposures and the ACS/WFC exposures for each satellite.
For And\,V,  the WFPC2 observations had two different pointings,
beyond the typical few-pixels dithering offsets.
The centers of the satellites are adopted from
\citet{mcconn2012}, and the $(\xi, \eta)$ coordinates
are the gnomonic projection of the equatorial coordinate system
about the center of the satellite.

\begin{deluxetable*}{cccr}
  \tablecaption{Properties of the Data Sets
    \label{tab:prop-data}}
\tablewidth{0pt}
\tablehead{
    \colhead{Camera} &
    \colhead{Epoch} &
    \colhead{Exposures} &
    \colhead{PID} 
}
\startdata
\multicolumn{4}{c}{\bf{And\,V}} \\
WFPC2    & 1999.86-2000.96 & $16\times1300$s (F450W) & 8272 \\
WFPC2    & 1999.86-2000.96 & $8\times1200$s (F555W) &  8272 \\ \\
ACS/WFC  & 2020.73 & $1\times60$s~;~$1\times900$s~;~$1\times950$s~;~$2\times980$s~;~$2\times985$s~;~$4\times1035$s (F606W) & 15902 \\ 
ACS/WFC  & 2020.73 & $1\times60$s~;~$1\times1080$s~;~$1\times1140$s~;~$2\times1148$s~;~$4\times1200$s~;~$3\times1260$s (F814W) & 15902 \\
\hline
\multicolumn{4}{c}{\bf{And\,VI}} \\
WFPC2    & 1999.81 & $16\times1300$s (F450W) & 8272 \\
WFPC2    & 1999.81 & $8\times1100$s (F555W) &  8272 \\ \\
ACS/WFC  & 2019.86 & $1\times60$s~;~$1\times900$s~;~$1\times950$s~;~$4\times980$s~;~$4\times1030$s (F606W) & 15902 \\ 
ACS/WFC  & 2019.86 & $1\times60$s~;~$1\times1080$s~;~$1\times1140$s~;~$4\times1200$s~;~$4\times1260$s (F814W) & 15902 \\ 
\enddata
\end{deluxetable*}

%\section{Data Processing \label{sec:proc}}

The detailed processing procedure for both WFPC2 and ACS/WFC images
has been described in \citet{casetti2024b}. Here, we will
briefly mention the main steps.

\subsection{WFPC2 \label{subsec:wfpc2}}
The WFPC2 standard-calibrated $_{-}$c0m.fits images
from MAST are 
corrected for cosmic-ray (CR) contamination using pairs of exposures
with offsets less than 1 pixel. We then use the 
{\it hst1pass} code, 2023 version \citep{and2000,and2022} to
obtain magnitudes and preliminary source $(x,y)$ centers in all exposures.
These preliminary centers are refined using the deep-learning (DL) model developed in
\citet{casetti2024a}. The DL model was developed for filters
F555W and F814W. Here, for filter F450W we will use the F555W model.
Likewise, an {\it hst1pass} ePSF library for F450W does not exist, thus
the ones for F555W were used. By using the DL model we aim to overcome
undersampling centering issues in the WFPC2 instrument, a procedure
that has been shown to be effective \citep{casetti2024a,casetti2024b}.

Once raw pixel positions are obtained, the following corrections are made:
the 34th-row correction \citep{and1999},
classic 3rd-order distortion correction \citep{and2003}, followed by the
more recent higher-order distortion corrections mapped by \citet{casetti2021}. 
As in the case of the centering process, there are no corrections for filter
F450W, therefore we used those developed for F555W.

We estimate the precision of the positions thus obtained,
via a transformation between a
reference exposure and a target exposure.
The transformation is a polynomial one that includes up to
3rd-order terms \citep[see e.g.,][]{casetti2021}. The
scatter of the residuals represents the errors in position
of both the reference and the target image. For 
well-measured stars, we obtain single-measurement standard
errors of 53 mpix for the planetary camera (PC) chip and 42 mpix
for the wide field (WF) chips, in filter F555W.
This corresponds to 2.4 mas for the
PC and 4.2 mas for the WF.
Positional errors for filter F450W are a few percent larger than those 
for F555W.

\subsection{ACS/WFC \label{subsec:acs}}

For the ACS/WFC images, we work with $_{-}$flt.fits files
from MAST that have undergone
the standard HST-pipeline calibration.
Detections, positions and magnitudes are obtained
using {\it hst1pass}. The positions obtained have undergone all
astrometric corrections implemented by {\it hst1pass} (2023 version),
including distortion and charge transfer efficiency correction.
More specifically,
the distortion is based on the work done in
\citet{kp2015} and \citet{kp2018}. 
Standard ePSF library files and distortion
corrections for each filter of ACS data exist, and
were used here.

As in Sec. \ref{subsec:wfpc2}, we perform polynomial
transformations between exposures taken at the same epoch and in the
same filter to assess the positional precision.
Single-measurement standard errors are calculated for well-measured stars.
We obtain $\sim 17$ mpix for both filters, corresponding to
0.85 mas.

\section{Proper Motion Calculations \label{sec:pms}}

Following the procedure in \citet{casetti2024b}, we treat
each chip of each camera as a separate unit, that is
each chip has its own individual $(x,y)$ system. 
Pixel coordinates, obtained
in Secs. \ref{subsec:wfpc2} and \ref{subsec:acs},
are converted to equatorial coordinates, $(\alpha,\delta)$,
using the WCS information
in the header of the image fits files\footnote{ 
We do not rely on the WCS coefficients being very precise;
they are not.
Subsequent transformations between chips essentially override these approximate ones.
However, from a practical standpoint, 
working in a common equatorial system facilitates
star matching and simplifies the overall process.}.
Equatorial coordinates for all detections on each chip
are then gnomonically projected
into $(\xi, \eta)$ standard coordinates assuming a common tangent point,
which is taken to be the position of
the center of the satellite. We use  
$(\alpha,\delta)$ = (17.57125,  47.62806) degrees for And\,V
and $(\alpha,\delta)$ = (357.94292,  24.58250) degrees for And\,VI
\citep{mcconn2012}.

%Given the geometry of the various exposures' overlap (Fig. \ref{fig:overlap}),
For each satellite, we construct two separate proper-motion catalogs,
corresponding to the two ACS chips.
As reference exposure, we adopt the
initial exposure of the 2020 ACS/WFC F814W data set.
The $(\xi, \eta)$ positions from all other chips/exposures are transformed 
into either of the chips of this exposure, using polynomial transformations
with up to 4th-order coefficients in each coordinate. 

As can be seen in Fig. \ref{fig:overlap}, the overlap between
early and late epochs is not ideal. In And\,V's case we could
not achieve a solution between the PC and ACS chip 2. This is due to the
small overlap and thus few reference stars to perform a reliable
solution. In And\,VI's case, more than half of one
WF chip of the WFPC2 is outside of the
ACS field.  Although a solution is obtained for this WF chip,
the area coverage, and hence the number of And\,VI stars in
the proper-motion determination, is reduced.

Proper motions are calculated using an iterative
least-squares procedure to refine both the polynomial coefficients relating
each chip/exposure into the reference exposure, and every object's proper motion;
the initial iteration assumes zero proper motions for all reference stars.
The reference stars that are used to perform the transformations
and compute the polynomial coefficients
are predominantly satellite stars. Thus the proper-motion system is relative,
and it is that of the satellite. We use only relatively bright,
well-measured stars in these transformations, with a typical faint limit
about a half magnitude below the horizontal branch of the satellite.

Proper motions\footnote{Throughout the paper, $\mu_{\alpha}$
is actually $\mu_{\alpha}~ cos~\delta$, and as units for
the proper motions we will specify either mas yr$^{-1}$ or $\mu$as yr$^{-1}$.}
are determined for all objects that have a minimum
of 10 separate position measurements and a minimum of 20 years time baseline.
The proper motion is computed as the slope of a simple linear fit
to each coordinate as a function of
time, removing the highest outlier if it deviates more than $2.5\sigma$ 
from the best-fit line, until no such outliers remain. 
Formal relative proper-motion uncertainties are calculated
from the scatter about this best-fit line. The specific time baselines
are 20.9 years for And\,V and 20.1 years for And\,VI.

\begin{figure*}
    \centering
    \includegraphics[scale=0.38,angle=-90]{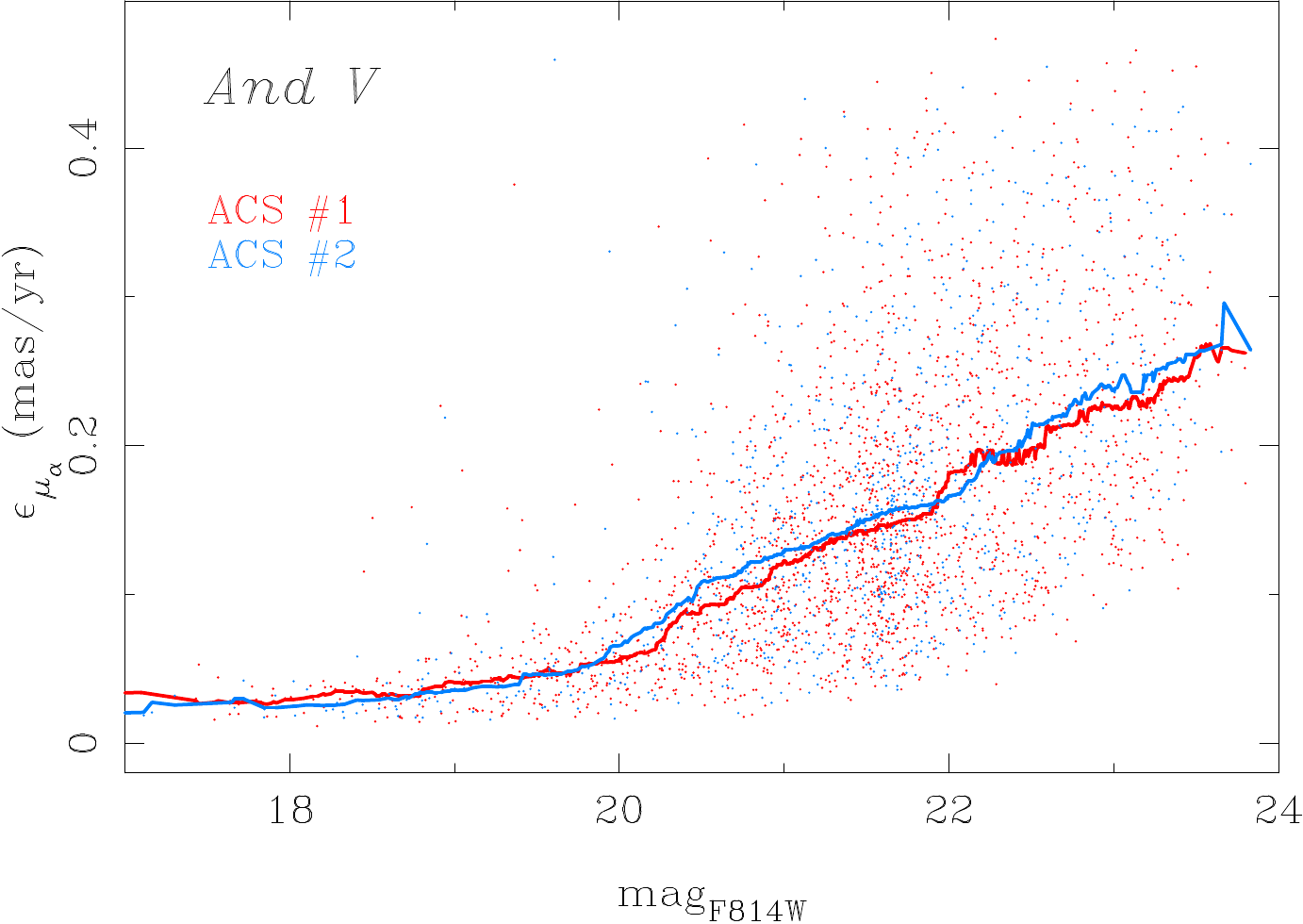}
    \includegraphics[scale=0.38,angle=-90]{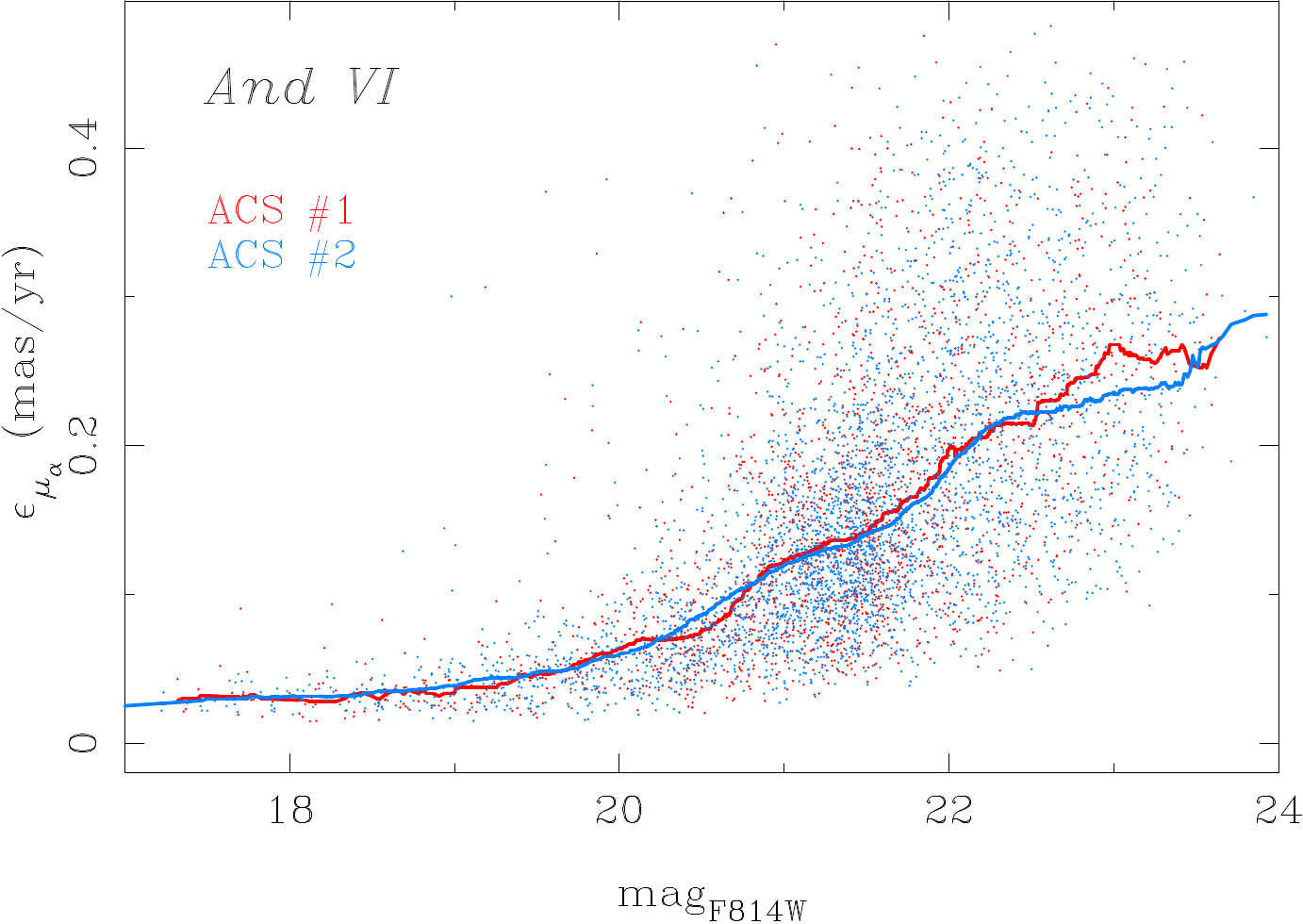}
    \caption{Proper-motion uncertainty as a function of
      instrumental magnitude in F814W for both ACS catalogs/chips and satellites. 
      A moving median (1-mag bin) is represented with a solid line for each sample.
      The horizontal branch of each system is
      at mag$_{F814W} \sim 21.5$, where a higher stellar density is apparent.}
    \label{fig:pmerr-mag}
\end{figure*}

In Figure  \ref{fig:pmerr-mag} we illustrate the run
of the proper-motion uncertainty in $\mu_{\alpha}$
with F814W instrumental magnitude.
Similar looking plots are obtained for $\mu_{\delta}$.
Well-measured stars have uncertainties between 25 and $50~\mu$as yr$^{-1}$.

\begin{figure*}
    \centering
    \includegraphics[scale=0.70,angle=0]{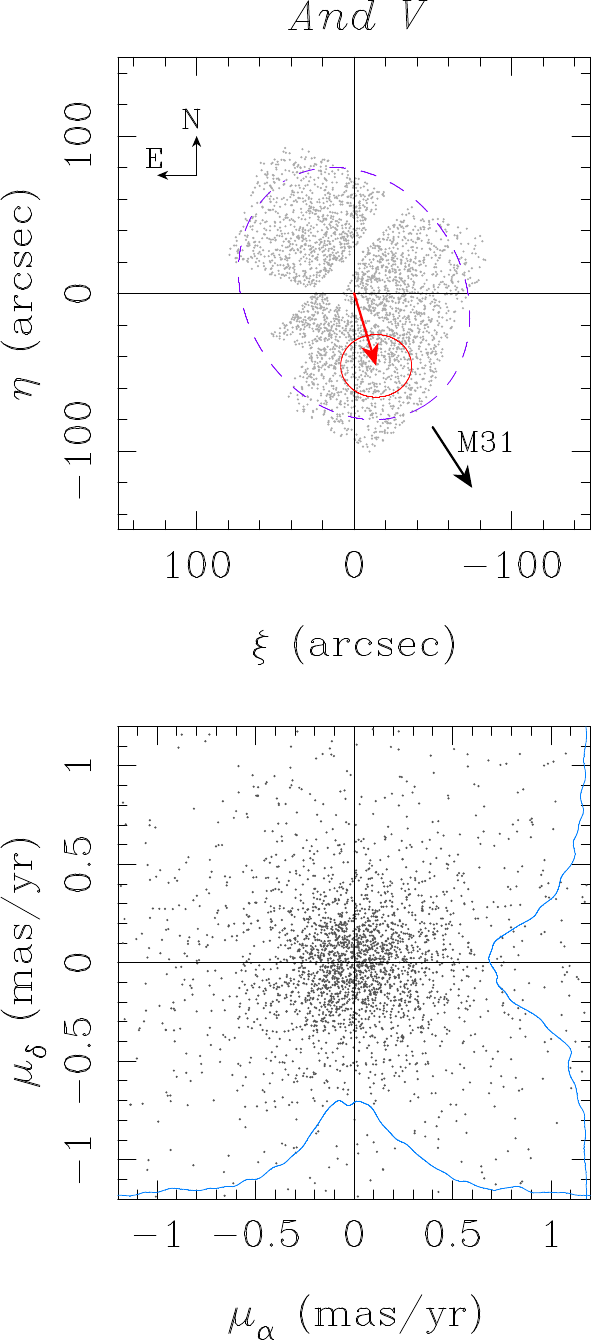}
    \includegraphics[scale=0.70,angle=0]{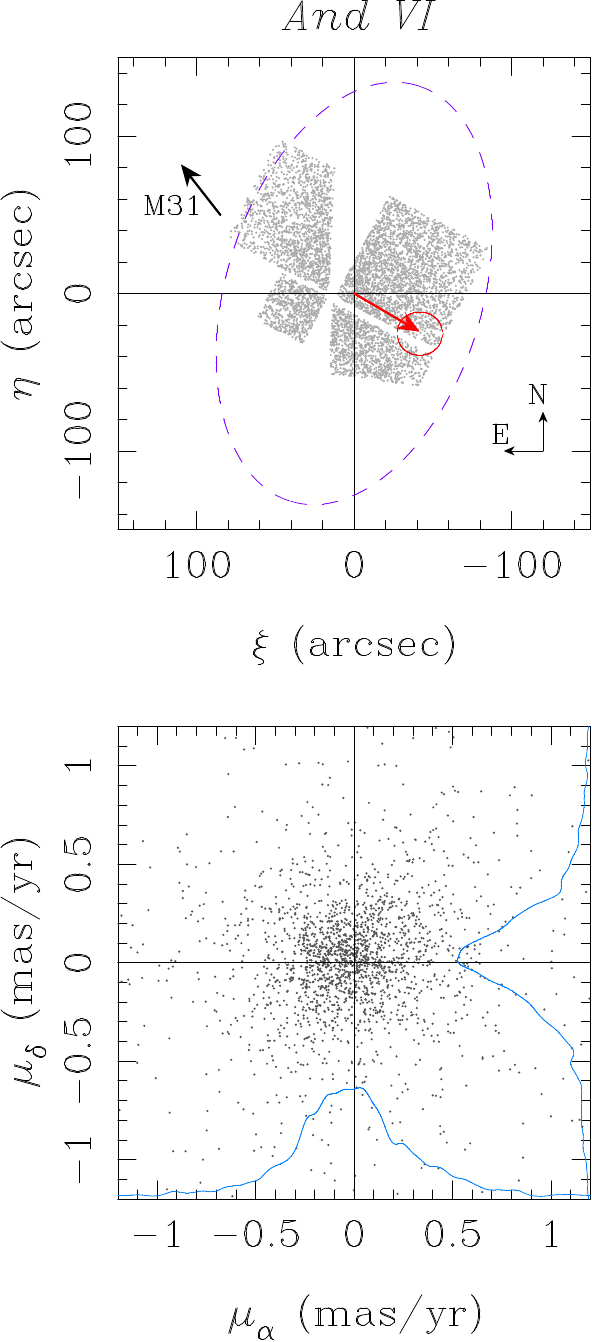}
    \caption{Proper-motion catalogs:
      spatial distribution (top), and relative proper motions
for stars with proper-motion uncertainties
$\le 0.5$ mas yr$^{-1}$ (bottom). In the top panels we mark the direction of the satellite
proper motion {\bf relative to that of M31} (red arrow) together with its error ellipse and 
the direction toward M31. 
The half-light ellipse, according to the parameters in \citep{mcconn2012}, is represented with 
a dashed line. In the bottom panels, the blue curves represent the proper-motion histograms in 
each coordinate.}
    \label{fig:vpd}
\end{figure*}

The sky distribution and the relative proper-motion distribution for each satellite
are shown in Figure \ref{fig:vpd}. Here we use only objects with 
combined proper-motion uncertainties
$\sqrt{(\epsilon_{\mu_{\alpha}}^2+\epsilon_{\mu_{\delta}}^2)} \le 0.5$
mas yr$^{-1}$.

\section{Absolute Proper Motions \label{sec:absolute}}

Although we have some {\it Gaia} Early Data Release 3 (EDR3) \citep{edr3}
stars measured in our proper-motion catalogs, the EDR3 proper-motion
errors are unfortunately too large to provide a useful
determination as was the case in \citet{casetti2024b}.
This is because stars in our catalogs reach only the faint end of
EDR3. Specifically, for And\,V we find nine EDR3 stars,
with eight in the ACS chip-1 catalog and one in the
ACS chip-2 catalog. For And\,VI we have only one EDR3 star per catalog.
In the case of And\,V ACS chip-1 catalog, the weighted average of the EDR3 stars
has uncertainties of between 0.1 to 0.2 mas yr$^{-1}$, while in the other cases,
uncertainties are 1 to 2 mas yr$^{-1}$ , thus being inadequate for our purpose.
Future {\it Gaia} Data Releases, with improved faint-end precision,
could potentially be used at a later time.

For now, we rely solely on background galaxies to determine the correction to
absolute proper motions. We use only the ACS F814W images to identify galaxies.
In a first pass we do an eye selection on a median-combined image.
In a second pass we use the sextractor classifier \citep{bertin2010} on each CR-cleaned image.
We obtain a class parameter for each image, which is then averaged and
subsequently used to add more galaxies at the faint end of the sample.
These sextractor-classified galaxies are also eye inspected.
We very carefully eliminate ``compromised'' galaxies
which were blended with a star or lying on diffraction spikes of bright stars.

%tg_begin
Once identified as such, proper motions of these zero-point reference galaxies
are extracted from the relative proper-motion catalogs described in the previous section.
Note that astrometric centers for galaxy images come directly from the hst1pass values
for ACS data, and from the DL-model for the WFPC2 data, just as for the stellar images.
The effect of the (im)precision of these centers on an individual galaxy's proper-motion 
measure is determined from the scatter of residuals in its proper-motion solution, and 
propagated into its proper-motion uncertainty estimate.
Thus, the expected larger random errors for galaxy images is explicitly included in the
analysis.
%tg_end

The final zero-point correction is a weighted average of all galaxies
\footnote{
It is felt that the alternative use of local zero-point corrections, made to stars
near each individual galaxy \citep[e.g.,][]{sohn2012,dinescu1997}, would not provide a superior correction given the
number of galaxies, their proper-motion uncertainities, and the small size of the
independently determined proper-motion catalogs, i.e., based on the footprint of
each WFPC2 detector.}
with total proper-motion values less than 2 mas yr$^{-1}$.
The weights are derived from the individual formal proper-motion errors of each galaxy. 
These zero-point corrections are listed in Table \ref{tab:abs-pm}
as $\mu^{cor}$
together with the number of galaxies used per ACS chip solution and satellite.

The systemic relative proper motion of the satellite is determined
as the mean motion of satellite members.
We limit this sample to stars with mag$_{F814W}$ between 17 and 22,
with combined proper-motion uncertainties
$\le 0.5$ mas yr$^{-1}$, and with total
proper motion  $\le 1.0$ mas yr$^{-1}$.
We also use color-magnitude information
obtained from the ACS exposures to eliminate photometric non-members.
The mean and its uncertainty are computed using probability plots
\citep{hamaker1978} with trimming of the extreme $10\%$
of both wings to eliminate the influence of outliers.
In Table \ref{tab:abs-pm} we list the relative mean
proper motion ($\mu^{rel}$) together with the number of satellite
stars used in each solution.

The absolute proper motion is the straight difference between
the mean relative proper motion and the zero point. These
values per chip solution and satellite are also listed in the
last two columns of Tab. \ref{tab:abs-pm}.

\begin{deluxetable*}{lrrrrrrrr}
  \tablecaption{Relative and Absolute Mean Proper Motions for Each ACS chip Solution
    \label{tab:abs-pm}}
\tablewidth{0pt}
\tablehead{
  \colhead{Cat.} &
    \colhead{$\mu_{\alpha}^{cor}$} &
    \colhead{$\mu_{\delta}^{cor}$} &
    \colhead{$N_g$} &
    \colhead{$\mu_{\alpha}^{rel}$} &
    \colhead{$\mu_{\delta}^{rel}$} &
    \colhead{$N_s$} &
    \colhead{$\mu_{\alpha}^{abs}$} &
    \colhead{$\mu_{\delta}^{abs}$} \\
    \colhead{} &
    \colhead{($\mu$as yr$^{-1}$)} &
    \colhead{($\mu$as yr$^{-1}$)} &
    \colhead{} &
    \colhead{($\mu$as yr$^{-1}$)} &
    \colhead{($\mu$as yr$^{-1}$)} &
    \colhead{} &
    \colhead{($\mu$as yr$^{-1}$)} &
    \colhead{($\mu$as yr$^{-1}$)} 
    }
\startdata
&&&& {\bf{And\,V}} &&&& \\
\#1 &  $-36\pm25$ & $75\pm22$ & 68 & $7\pm 8$&  $-7\pm 8$ & 891 & $43\pm 26$&  $-82\pm23$ \\
\#2 &  $16\pm35$ & $50\pm32$ & 26 & $6\pm 16$&  $-7\pm 12$ & 403 & $-10\pm 38$&  $-57\pm34$ \\
\hline
\hline
&&&& {\bf{And\,VI}} &&&& \\
\#1 &  $-27\pm12$ & $72\pm11$ & 67 & $-18\pm 8$&  $23\pm 7$ & 1015 & $9\pm 14$&  $-49\pm13$ \\
\#2 &  $34\pm25$ & $69\pm21$ & 71 & $-4\pm 7$&  $6\pm 6$ & 1810 & $-38\pm 26$&  $-63\pm22$ \\
\enddata
\end{deluxetable*}

For both satellites the ACS chip-1 solution appears formally better than
the chip-2 solution. This is because chip-1 solution includes most
or all of the PC of WFPC2 (see Fig. \ref{fig:overlap}) which has a
better resolution than the WF chips.

Finally, the adopted
absolute proper motion is taken to be the
error-weighted average of the two ACS chip solutions.
For And\,V we obtain an absolute proper motion of
$(\mu_{\alpha} , \mu_{\delta}) = (26.1\pm21.5, -74.2\pm19.1)~\mu$as yr$^{-1}$.
For And\,VI we obtain an absolute proper motion of
$(\mu_{\alpha} , \mu_{\delta}) = (-1.6\pm12.3, -52.6\pm11.2)~\mu$as yr$^{-1}$.

Overall, the And\,VI determination is better than And\,V's
for the following reasons: 1)
And\,VI is a brighter, richer satellite than And\,V,
2) there are more background
galaxies in And\,VI than in And\,V (see Tab. \ref{tab:abs-pm},
Fig. \ref{fig:overlap}) and 3)
the entire PC chip overlapped with one single ACS chip
in And\,VI's case.

\section{Orbit Analysis \label{sec:orb-ana}}

\subsection{Space Velocities and Orbital Poles \label{subsec:vel-poles}}

The space position and velocity of each satellite are calculated
by combining our derived proper motion with
other observed parameters. The methodology is generally similar
to that adopted in \citet{casetti2024b}
for the orbital analysis of And\,III.

We use RR Lyrae-based heliocentric distances for M31
($776_{-21}^{+22}\,\mathrm{kpc}$), And\,V ($759\pm 21\,\mathrm{kpc}$),
and And\,VI ($832\pm23\,\mathrm{kpc}$) by \citet{savino2022}.
Line-of-sight velocities are 
$-300.1\pm3.9\:\mathrm{km}\,\mathrm{s}^{-1}$ \citep{mcconn2012} for M31
and 
$-397.3\pm1.5\:\mathrm{km}\,\mathrm{s}^{-1}$,
and $-339.8\pm1.9\:\mathrm{km}\,\mathrm{s}^{-1}$
\citep[][their Table 5]{Collins2013},
for And\,V and And\,VI respectively.

The value we adopt for M31's proper motion strongly
influences the resulting kinematics, in particular for And\,V.
We test two PM estimates: the HST + Sats value used
in \cite{sohn2020} of $(\mu_{\alpha}, \mu_{\delta})=(34.3\pm8.4, -20.2\pm7.8)\:\mu\mathrm{as}\,\mathrm{yr}^{-1}$,
as well as its weighted average with the EDR3 proper motion
by \citet{salomon2021}.
The latter $(\mu_{\alpha}, \mu_{\delta})=(40.1\pm6.6, -28.3\pm5.6)\:\mu\mathrm{as}\,\mathrm{yr}^{-1}$,
calculated in \citet{ps2021}, arguably represents our
current best estimate of M31's systemic proper motion using published data.

\begin{figure}
    \centering
    \includegraphics[scale=0.60,angle=0]{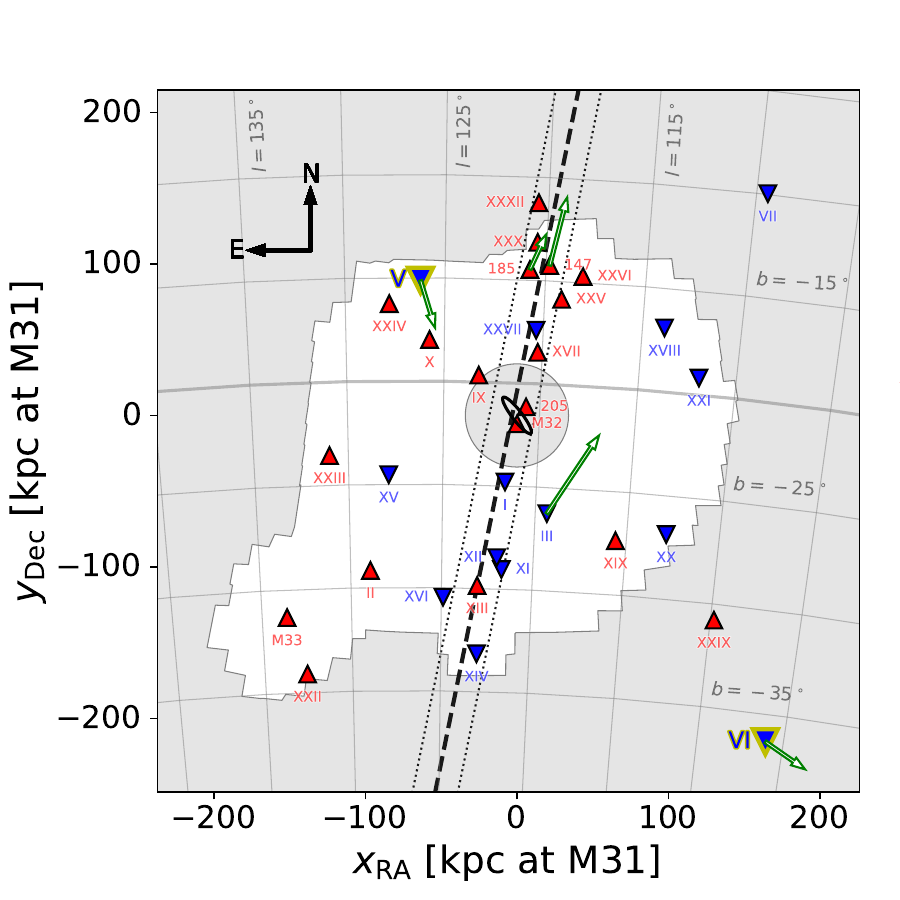}
    \caption{Spatial distribution and motions of satellite galaxies around M31.
      The position and orientation of the M31 disk is indicated with a black ellipse,
      the white area indicates the extent of the PAndAS footprint.
      Red upward triangles mark dwarf galaxies with receding line-of-sight velocities
      relative to that of M31, blue downward triangles those with approaching velocity.
      The on-sky velocities relative to M31 for satellite galaxies with measured
      proper motions are marked with green arrows, with And\,V and VI,
      whose proper motions are presented in this work, highlighted with yellow outlines.
      The Great Plane of Andromeda (GPoA) and its root-mean-square width are
      shown with dashed and dotted lines, respectively.}
    \label{fig:mapsat}
\end{figure}

The on-sky distribution of satellite galaxies around M31 is shown
in Figure~\ref{fig:mapsat},
together with the directions of motion
relative to M31 calculated from the measured proper motions,
assuming the HST + Sats + EDR3 proper motion for M31.

Neither of the two galaxies is part of the Great Plane 
of Andromeda (GPoA), M31's satellite plane,
with And\,V being at least $43.5^\circ$
and And\,VI at least $47.5^\circ$ offset from the plane.
Consequently, they can not orbit along this structure.
However, both And\,V and And\,VI are consistent with 
aligning to within a few degrees with the galactic disk of M31.
Both are part of the M31 disk plane satellite galaxies
in \citet{pkj2013}.
%Not sure about this phrase: an observer located in M31's disk??
An observer with the equivalent of a Sun-like vantage-point in M31 would most
likely not observe these galaxies,
as they would be hidden behind the galactic disk of M31.

\begin{figure}
    \centering
    \includegraphics[scale=0.40,angle=0]{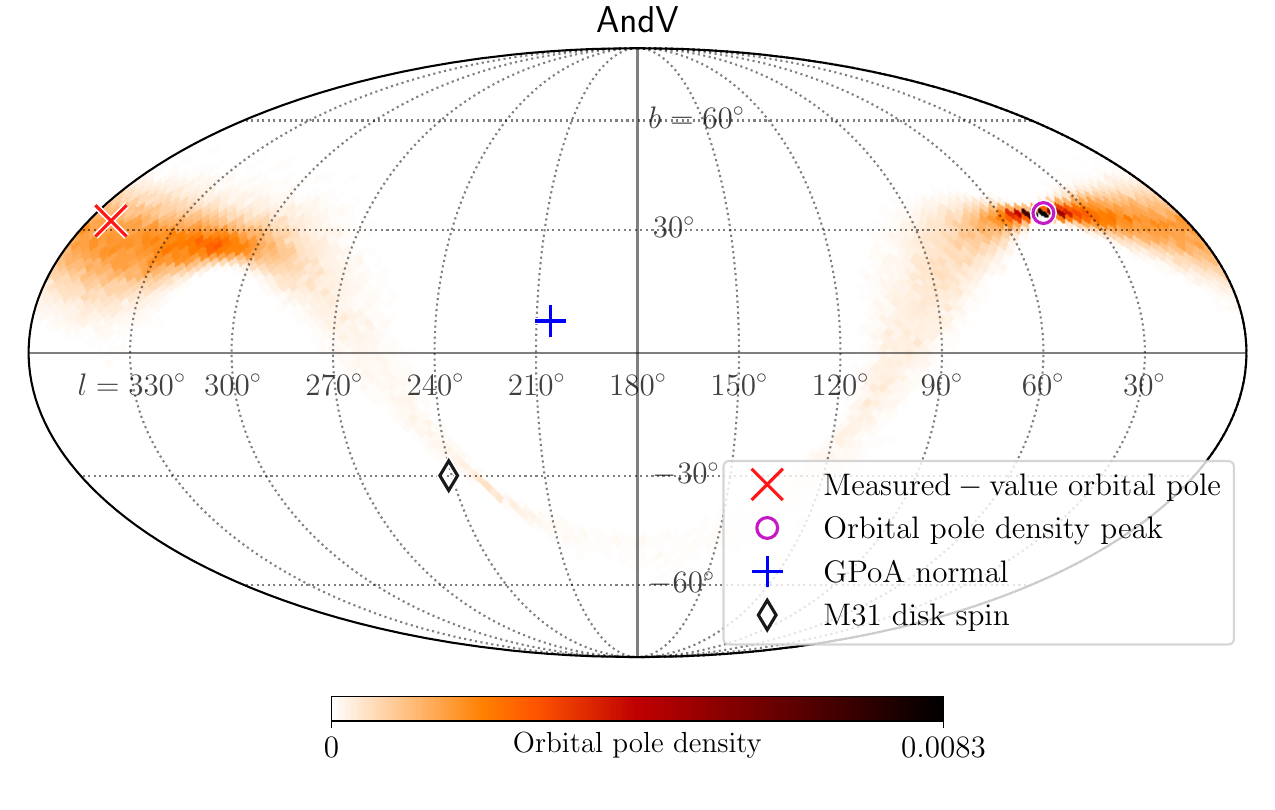}
    \includegraphics[scale=0.40,angle=0]{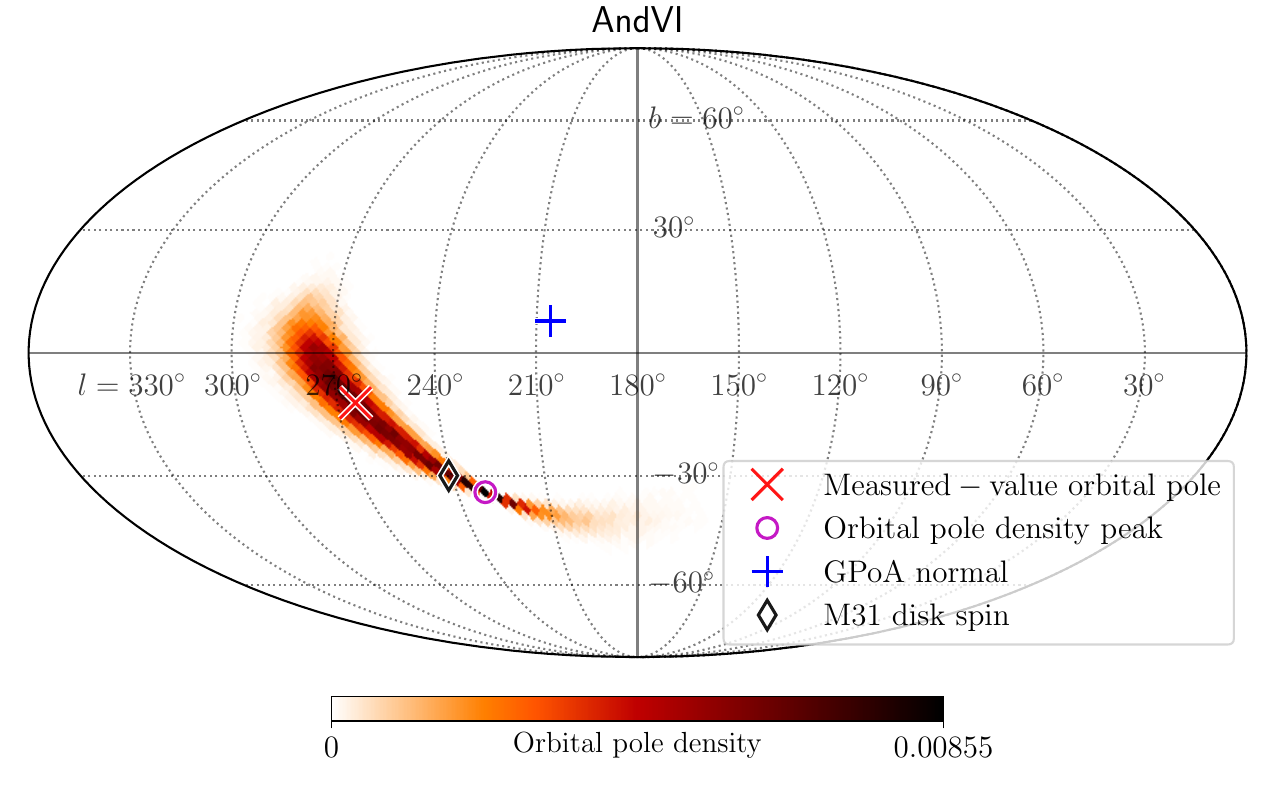}
    \caption{Distribution of calculated orbital poles of And\,V (top) and VI (bottom)
      relative to M31 in Galactic coordinates from $10^5$ Monte Carlo realizations.
      The orbital pole corresponding to the measured-value position and velocity, i.e.,
      ignoring uncertainties, is shown as a red cross. The peak in the density distribution
      when including the uncertainties is shown as a magenta circle.
      The black diamond indicates the spin of the galactic disk of M31,
      while the blue plus sign indicates the normal vector to the GPoA
      aligned with its direction of co-rotation. And\,VI is consistent with
      co-orbiting along the galactic disk spin of M31, while And\,V
      is counter-orbiting and less well aligned.}
    \label{fig:orbpoles}
\end{figure}

To investigate whether the satellite galaxies move out of
the M31 galactic disk plane or, alternatively, orbit along the
disk, we present in Figure~\ref{fig:orbpoles} the
orbital pole distributions 
%staying forever hidden for an M31-based observer,
%Figure~ \ref{fig:orbpoles} shows the orbital pole distribution of the two galaxies
for $10^5$ Monte-Carlo realizations drawing from the
respective uncertainties. We adopt here the HST+Sats+EDR3
proper motion of M31 as a reference frame.
Both galaxies are --- in general --- consistent
with moving along the M31 equatorial plane,
within the orbital uncertainties.
And\,VI is better aligned, compared to And\,V.
The two satellites also orbit in opposite senses:
And\,V counter-orbits with respect
to M31's galactic spin, while And\,VI co-orbits.

Specifically, for And\,VI, its 'measured-value' proper motion, distance and
line-of-sight velocity
imply an orbital (mis)alignment of $28^\circ$ with respect to the plane
defined by the galactic disk of M31, moving in a direction co-orbiting with
M31's galactic spin. 
Alignment to within $11^\circ$ is consistent with 1$\sigma$ measurement
uncertainties in these quantities, while alignment within $2^\circ$
is consistent with 2$\sigma$ uncertainties.
A median alignment angle of $(30 \pm 15) ^\circ$ is found when considering
all Monte-Carlo realizations.

For And\,V, its 'measured-value' properties imply an orbital (mis)alignment 
of $58^\circ$ with the counter-orbiting M31 spin direction. Considering the 
Monte-Carlo realizations a median alignment angle of $(61 \pm 35) ^\circ$ is 
found and an alignment to within $22^\circ$ ($5^\circ$) is consistent within 
1$\sigma$ (2$\sigma$).

\subsection{Properties of the Integrated Orbits \label{subsec:orb-prop}}

\begin{deluxetable*}{crrrrrrrr}
\tablecaption{Orbital Parameters
	\label{tab:orbits}}
\tablewidth{0pt}
\tablehead{
	\colhead{M31 PM} &
	\colhead{$M_{\mathrm{vir,M31}}$} &
	\colhead{$f_{\mathrm{peri}}$\tablenotemark{a}} &
	\colhead{$t_{\mathrm{peri}}$\tablenotemark{b}} &
	\colhead{$r_{\mathrm{peri}}$\tablenotemark{c}} &
	\colhead{$f_{\mathrm{apo}}$\tablenotemark{a}} &
	\colhead{$t_{\mathrm{apo}}$\tablenotemark{b}} &
	\colhead{$r_{\mathrm{apo}}$\tablenotemark{c}} &
	\colhead{$e$\tablenotemark{d}} \\
	\colhead{} &
	\colhead{($\times10^{12}\,M_{\odot}$)} &
	\colhead{(\%)} &
	\colhead{(Gyr)} &
	\colhead{(kpc)} &
	\colhead{(\%)} &
	\colhead{(Gyr)} &
	\colhead{(kpc)}
}
\startdata
&&&& {\bf And\,V}& & &\\
HST+Sats & 1.5 & 64 & 4.6 $[2.2, 7.2]$ & 49 $[30, 81]$ & 78 & 2.1 $[1.0, 5.0]$ & 292 $[166, 551]$ & 0.71 $[0.52, 0.87]$\\
       " & 2.0 & 82 & 3.1 $[1.9, 6.2]$ & 46 $[29, 78]$ & 91 & 1.4 $[0.8, 3.7]$ & 231 $[156, 482]$ & 0.66 $[0.47, 0.84]$\\
\hline
HST+Sats+EDR3 & 1.5 & 76 & 3.4 $[1.9, 6.7]$ & 42 $[29, 78]$ & 87 & 1.5 $[0.8, 4.3]$ & 232 $[149, 493]$ & 0.69 $[0.44, 0.85]$\\
            " & 2.0 & 89 & 2.5 $[1.6, 5.1]$ & 39 $[27, 72]$ & 95 & 1.0 $[0.6, 2.9]$ & 196 $[143, 393]$ & 0.66 $[0.45, 0.83]$\\
\hline
\hline
&&&& {\bf And\,VI}& & &\\
HST+Sats & 1.5 & 99 & 1.8 $[1.3, 2.4]$ & 100 $[71, 170]$ & 89 & 5.2 $[4.6, 6.8]$ & 334 $[289, 486]$ & 0.54 $[0.42, 0.69]$\\
       " & 2.0 & 100 & 1.7 $[1.3, 2.2]$ & 87 $[64, 147]$ & 97 & 4.4 $[4.1, 5.7]$ & 319 $[288, 435]$ & 0.57 $[0.43, 0.69]$\\
\hline
HST+Sats+EDR3 & 1.5 & 99 & 2.0 $[1.2, 2.6]$ & 143 $[84, 215]$ & 87 & 5.7 $[4.7, 7.2]$ & 333 $[287, 479]$ & 0.4 $[0.33, 0.65]$\\
            " & 2.0 & 99 & 1.9 $[1.3, 2.4]$ & 122 $[78, 198]$ & 96 & 4.8 $[4.2, 6.1]$ & 315 $[287, 449]$ & 0.44 $[0.33, 0.64]$\\
\enddata
\tablecomments{Orbital parameters after integrating backward for 10 Gyr. Uncertainties are included as $[15.9, 84.1]$ percentiles.
  %Parameters denoted with a dash represent cases lacking a last pericentric or apocentric passage within the last 10 Gyr.
}
\tablenotetext{a}{Fraction of orbits that achieved peri/apocentric passage over the last 10 Gyr.}
\tablenotetext{b}{Lookback time to last peri/apocentric passage.}
\tablenotetext{c}{Distance from M31 at the last peri/apocentric passage.}
\tablenotetext{d}{Eccentricity of the orbit.}
\end{deluxetable*}

Orbital integrations for the two satellites are made based on an exploration
of two different mass models for M31's potential and two different absolute
proper-motion determinations of M31.
For M31's potential, 
we use the models from \citet{patel17} (their Table 2): a high-mass M31
with a virial mass of $M_{\mathrm{vir}}=2\times10^{12}\,M_{\odot}$ and
a low-mass M31 with $M_{\mathrm{vir}}=1.5\times10^{12}\,M_{\odot}$.
Both potentials also contain a central Hernquist bulge and Miyamoto-Nagai disc
to represent the baryonic galaxy. We estimate the effects of
Chandrasekhar dynamical friction for a Plummer sphere with
$M=2\times10^9\,M_{\odot}$ and $R=0.7\,\mathrm{kpc}$ (And\,V),
and $M=4\times10^9\,M_{\odot}$ and $R=1.0\,\mathrm{kpc}$ (And\,VI).
The strength of the dynamical friction implemented serves
as an upper bound, although we find that modifying this
component does not have a large impact on the resulting orbits.

Observational errors are accounted for by propagating
uncertainties for distances, line-of-sight velocities,
and proper motions for M31, And\,V, and And\,VI in a Monte Carlo fashion.
We generate 1000 initial conditions for backwards integration for
each satellite, adopting random offsets from the measured-value parameters
that are consistent with their stated uncertainties.
We integrate each satellite and its Monte Carlo realizations
backward for a period of 5 Gyr with M31's position fixed
at the origin throughout, although in some cases we extend this
to 10 Gyr for the purposes of obtaining reliable orbital parameters.

In the following discussion we adopt the HST + Sats + EDR3 proper motion
alongside the high-mass M31 potential as the {\bf fiducial model},
with other cases stated specifically.
Due to errors propagated in both distance and
proper motions for M31 and both satellites,
analytic orbits show considerable in their orbital properties
which are summarized in Table ~\ref{tab:orbits}.

When plotting orbits, space positions and velocities are
described by adopting the commonly used M31-centric
coordinate system as follows. Observed parameters are first
transformed into a Cartesian Galactocentric frame,
then translated into M31's rest frame. Finally, we apply
a rotation such that the X-Y plane is aligned with M31's
galactic disc, while the X-axis points away from the
Milky Way's direction. This is functionally equivalent to
the frame adopted in \citet{sohn2020}, although we instead adopt
updated distances from \citet{savino2022}.

\begin{figure}
%    \centering
    \includegraphics[scale=0.35,angle=0]{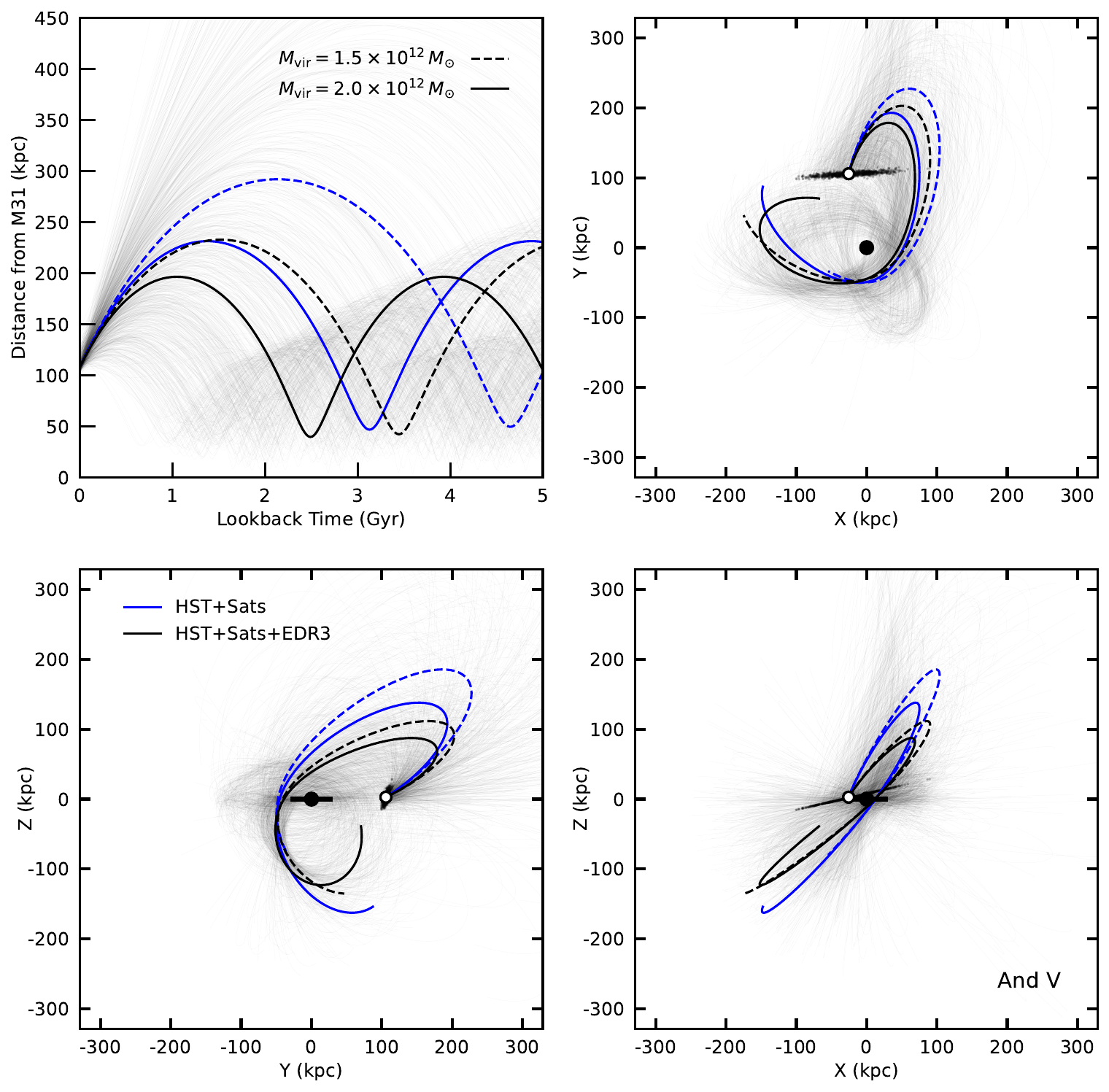}
    \caption{Backward orbit integrations of And\,V
      around M31 over the last 5 Gyr for a high-mass ($M_{\mathrm{vir}}=2\times10^{12}\,M_{\odot}$)
      and low-mass ($M_{\mathrm{vir}}=1.5\times10^{12}\,M_{\odot}$) M31 model,
      in solid and dashed lines respectively.
      The blue orbits assume the HST+Sats M31 proper motion,
      while the black orbits adopt the fiducial choice of a
      weighted average with M31's EDR3 proper motion.
      Andromeda's disc lies along the X-Y plane, while the Milky Way
      lies towards the negative X-axis.
      Present-time positions and orbital trajectories are also shown
      for Monte Carlo realizations of And\,V in the fiducial model (high-mass M31 and HST+Sats+EDR3 proper motion).
      %not sure what this means
      Note that the present-time position distribution (top, right panel) has a width orthogonal
      to the line-of-sight direction due to contributions from uncertainties in M31's distance.
    }
    \label{fig:orbits-and5}
\end{figure}

\begin{figure}
%    \centering
    \includegraphics[scale=0.35,angle=0]{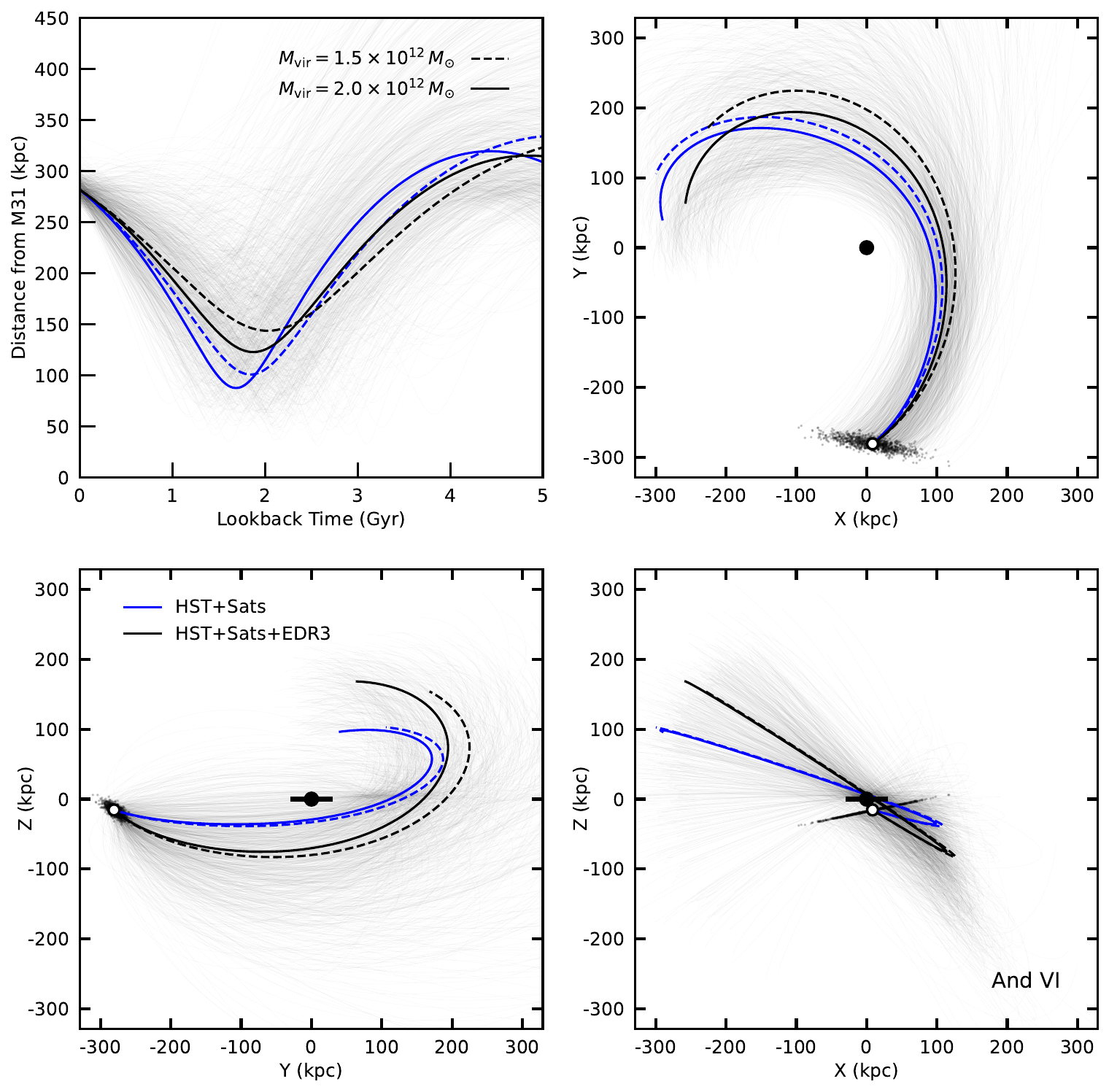}
    \caption{Same an in Fig. \ref{fig:orbits-and5}, only for And\,VI.}
    \label{fig:orbits-and6}
\end{figure}

In all four realizations, And\,V is bound and currently
approaching pericenter, performing its next closest approach
in around 0.5 Gyr following a moderately eccentric orbit ($e=0.66^{+0.16}_{-0.22}$)
as seen in Figure~\ref{fig:orbits-and5} and Tab. \ref{tab:orbits}.
It is expected to approach M31 to within $r_\mathrm{peri} = 40$ to $50\,\mathrm{kpc}$.
However, the timing of its previous pericentric passage
(upper left panel of Fig.~\ref{fig:orbits-and5})
is heavily affected by both measurement uncertainties and the choice of model,
ranging between $\left<t_{\mathrm{peri}}\right>=2.5\,\mathrm{Gyr}$ --- for the fiducial model ---
and $4.7\,\mathrm{Gyr}$ --- for the HST+Sats proper motion and low-mass M31 potential.
The remaining panels in Fig.~\ref{fig:orbits-and5}
show the orbit of And\,V over the last 5 Gyr
in 2D projections of the previously defined Cartesian coordinate system.
In the fiducial model, $5\%$ of Monte Carlo realizations are unbound
(i.e., never reaching a previous apocenter),
but a majority of orbits are contained within a M31-centric distance of $300\,\mathrm{kpc}$.

The orbital characteristics of And\,VI are far better constrained,
due to comparative improvements in both distance and proper motion uncertainties.
Its orbit is shown in Figure~\ref{fig:orbits-and6}.
And\,VI is currently approaching
apocenter after a single pericentric passage
around $1.8\pm0.5\,\mathrm{Gyr}$ ago,
and will reach this point in around $1-2\,\mathrm{Gyr}$.
Its orbit is more circular than And\,V's with $e=0.46^{+0.19}_{-0.10}$,
well-bound ($f_{\mathrm{apo}}=0.96$, see Tab. \ref{tab:orbits}),
and consistent within uncertainties
with being co-planar with the galactic disc of M31.
At closest approach it is expected to remain beyond a distance
$r_\mathrm{peri} = 90$ to $140\,\mathrm{kpc}$ from M31.
Notably, the orbit calculations confirm the findings based on the orbital poles:
And\,VI orbits in the same rotational sense as the disc spin,
while And\,V counter-orbits. We also find that for both galaxies
the line-of-sight velocity is indicative of their orbital direction.
We point out that the choice of M31 mass
and proper motion does not strongly change the expected
orbital characteristics, with each scenario demonstrating
a last pericentric passage in $\left<t_{\mathrm{peri}}\right>=1.5-2\,\mathrm{Gyr}$
and a previous apocenter around $\left<t_{\mathrm{apo}}\right>=4-6\,\mathrm{Gyr}$
at a distance of $300-350\,\mathrm{kpc}$.

\begin{figure}
    \centering
    \includegraphics[scale=0.28,angle=0]{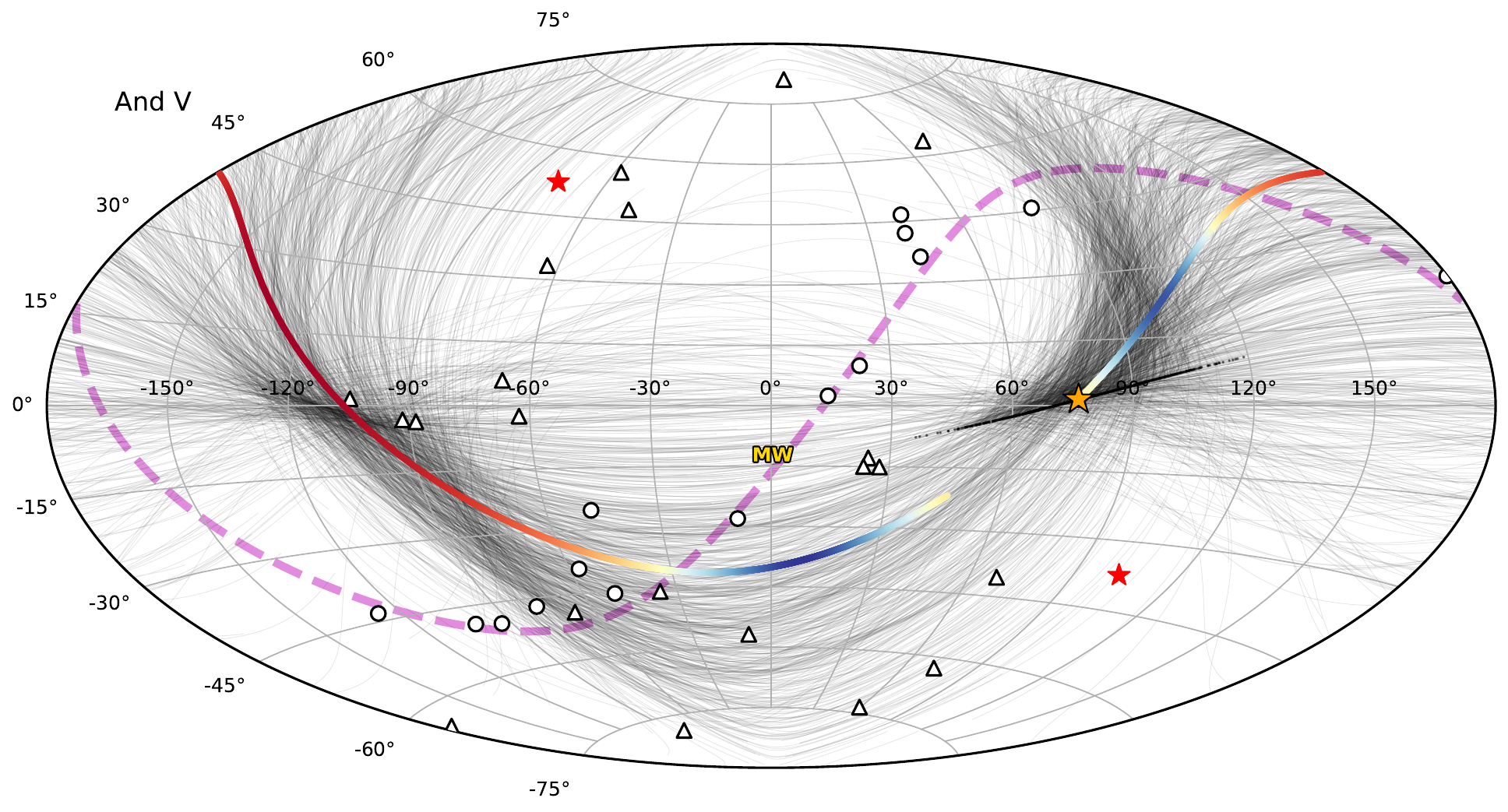}
    \includegraphics[scale=0.28,angle=0]{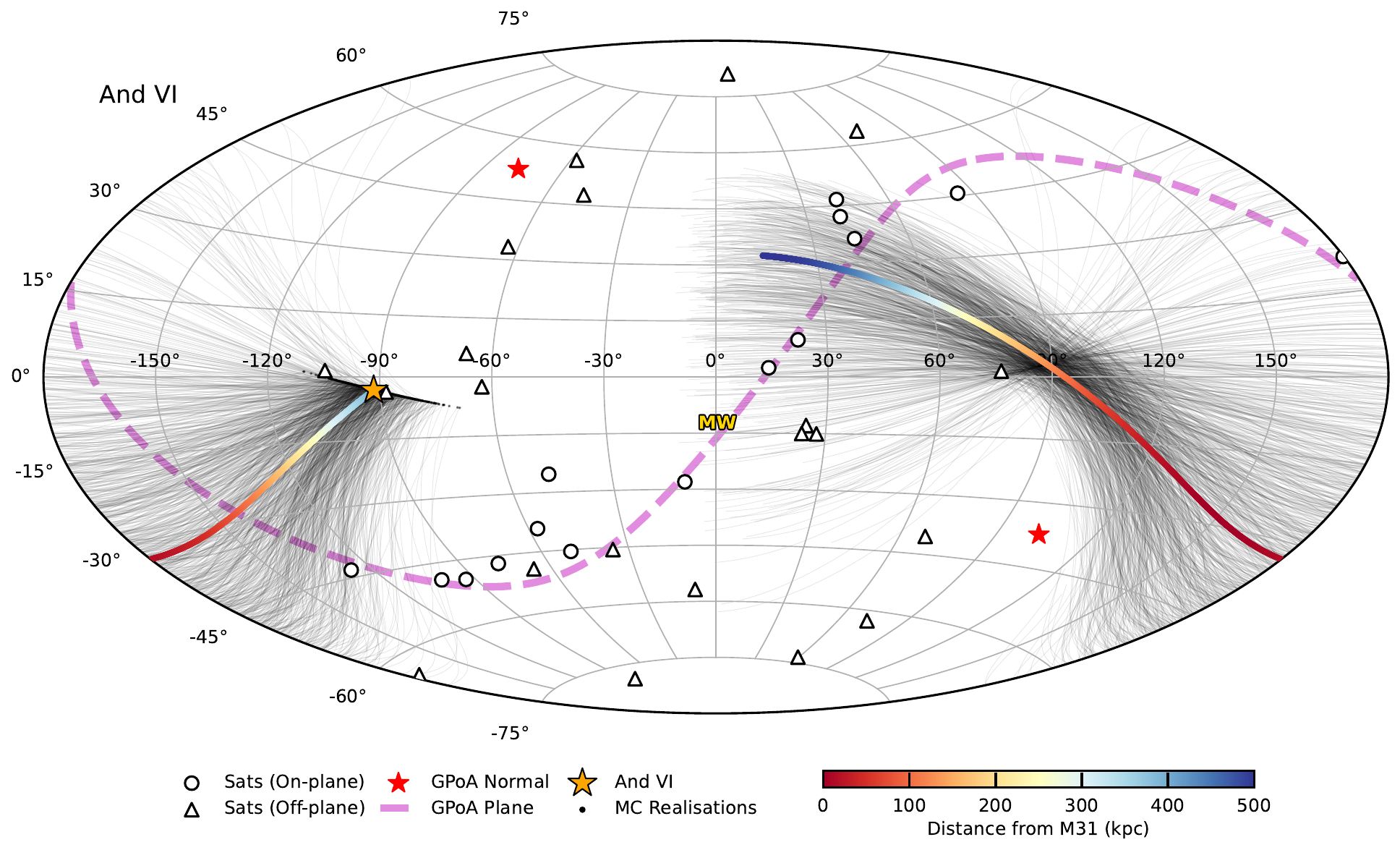}
    \caption{Sky projection of the M31 satellite system as seen from M31.
      The M31 disc is aligned with the equator, and the Milky Way lies along $\phi=0^\circ$.
      The yellow star represents the present-time position of And\,V (top) and And\,VI (bottom),
      while the thick line represents their most likely orbits coloured by M31-centric distance.
      The thin lines plot 1000 Monte Carlo realizations of the orbit,
      with each realization's present-time position marked with a small black dot.
      The other M31 satellites and the GPoA are also shown for reference.
      }
    \label{fig:sky-maps}
\end{figure}

In Figure \ref{fig:sky-maps} we show the past orbital tracks from our Monte-Carlo 
realizations in sky-projections as seen from the center of M31. Here, the galactic 
disk of M31 is oriented along the equator of the projection. The plots demonstrate 
that both And\,V and And\,VI have realizations that align very closely with this 
disk plane. Their opposite orbital directions are also apparent, as is the considerably 
less well constrained orbital orientation of And\,V compared to And\,VI. A majority 
of the known M31 satellites currently reside in the hemisphere facing the 
Milky Way \citep{savino2022}. In view of this considerable lopsidedness that 
challenges cosmological expectations \citep{Kanehisa2025}, it is interesting to 
note that both And\,V  and And\,VI had their last pericenters -- where their 
orbital motion is fastest and they thus spend the least time -- on the far side of M31.

\begin{figure}
    \centering
    \includegraphics[scale=0.60,angle=0]{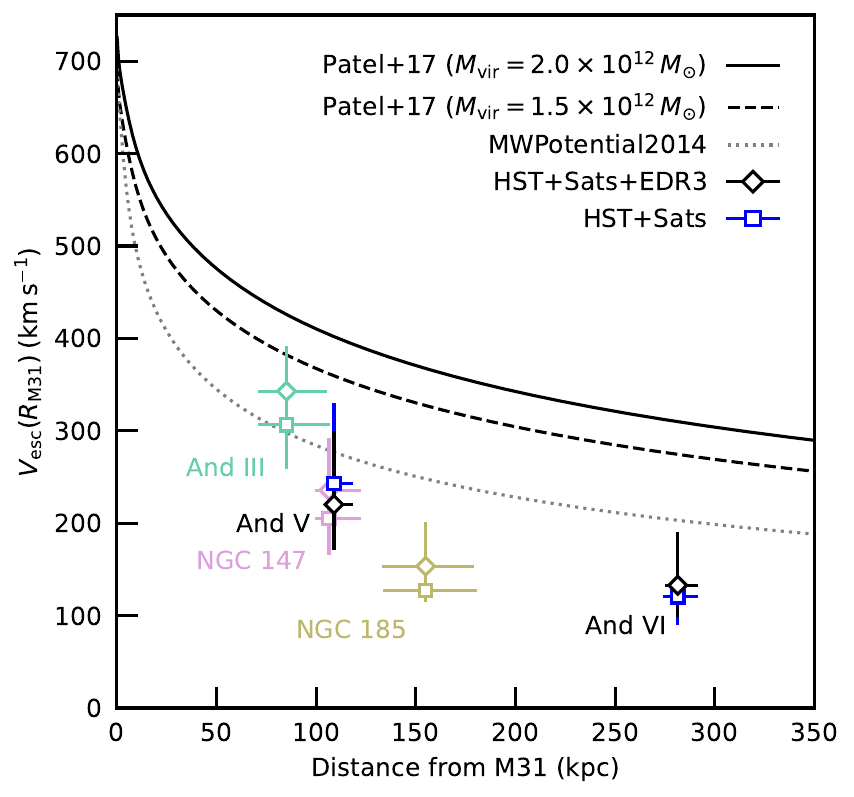}
    \caption{The M31-centric distance and total velocity of And\,V and VI using the
      HST+Sats M31 proper motion (blue square) and its weighted average with EDR3 (black diamond).
      Escape velocity curves of the two M31 mass models adopted in this paper are shown
      as solid and dashed lines, while a Milky Way-like potential is displayed for reference (grey dotted line).
      Results for NGC147, NGC185, and And\,III, three other M31 satellites with published
      proper motions, are also shown in pink, yellow, and teal respectively.
    }
    \label{fig:escvel}
\end{figure}

The expected energy of both satellites is bound, and lies well below
the escape velocity curve $v_{\mathrm{esc}}=\sqrt{-2\,\phi(r)}$ ---
where $\phi(r)$ is the combined gravitational potential of the
dark halo and baryonic galaxy component --- for all
four model combinations tested.
This is shown in Figure~\ref{fig:escvel}.

Indeed, with the exception of the HST + Sats proper motion with And\,V,
both satellites also lie fully below escape velocities
for a Milky Way-mass potential.
%As such, both satellites do not further constrain M31's mass,
%a feat only meaningfully performed among M31 satellites with
%published proper motions by And\,III \citep{casetti2024b}.
As such, both satellites do not further constrain M31's mass
directly under the assumption they are bound, a feat only
meaningfully performed among the M31 satellites with published
proper motions by And\,III \citep{casetti2024b}. However,
And\,V and VI's new proper motions can nevertheless be
beneficial for deriving mass constraints through
more sophisticated approaches using satellite phenomenology
\citep[e.g.][]{Patel2023}.

\section{Summary \label{sec:sum}}

We present the proper-motion measurements of two M31 satellites,
And\,V and And\,VI, located far away from the GPoA,
but close to the disk plane of M31. And\,VI is also
the farthest satellite from M31 ($\sim 280$~kpc)
among all six satellites with measured proper motions.

The measurements are based on
HST images taken 20 years apart with instruments that have been
thoroughly astrometrically calibrated. The correction to absolute
proper motion is given by 94 background galaxies for And\,V and
138 for And\,VI. While we have identified {\it Gaia}~EDR3
stars in our samples with as many as nine for And\,V and two for
And\,VI, these are currently unhelpful due to their large
EDR3 proper-motion errors. However, a future {\it Gaia}
data release with improved errors by as much as a
factor of two may be used to further pinpoint the
correction to absolute proper motion and compare
with the result given by background galaxies. Thus, having
relative proper motions of {\it Gaia}~EDR3 stars in our
catalog is another strength of this study.

Proper motions combined with line-of-sight velocities and
updated RR-Lyrae-based distances \citep{savino2022}
are used to calculate
the orbits of these two satellites around M31.
We explore two M31-mass models and two determinations
of M31's proper motion. We also perform a Monte-Carlo
analysis of the orbits considering
the uncertainties in the observed quantities.
We find that And\,VI closely aligns with the disk of M31
co-rotating with it, at moderate orbit eccentricity and
approaching its apocenter. The dwarf galaxy will remain
outside of $\sim 90$ kpc from M31, thus experiencing 
small tidal effects compared to the other satellites with known orbits.
\citet{Pickett2025} determine that the mass profile
of And\,VI falls in the cuspy regime, i.e., its central
density is large compared to that of other M31 satellites
(see Fig. 9 in \citet{Pickett2025}). They argue
that this is due to a weak tidal field acting on And\,VI
compared to the other satellites. Our orbit determination tends to
confirm this conjecture.

And\,V's orbit is less well-constrained
compared to And\,VI's due to both measurement uncertainties
and sensitivity to M31's proper motion. And\,V's orbit is
less well-aligned with M31's disk compared to that of And\,VI.
It is counter-orbiting the disk and currently moving toward pericenter.
Both satellites are well bound to M31, and thus do not immediatly
serve to further constrain the mass of M31.

Finally, we note that both And\,V  and And\,VI had
their last pericenters -- where they 
spend the least time -- on the far side of M31.
This may help better understand the lopsidedness
of the M31 satellite system where the known satellites
reside in the hemisphere facing the MW \citep{savino2022} ---
an unlikely feature in cosmological
simulations \citep{Kanehisa2025}.

\newpage
\acknowledgments
This work was supported by program HST-AR-17029
provided by NASA through a grant from Space Telescope
Science Institute, which is operated by the
Association of Universities for Research in Astronomy, Inc.
MSP acknowledges funding via a Leibniz-Junior Research Group (project number J94/2020).

This study has made use of data from the European Space Agency
(ESA) mission {\it Gaia} 
(\url{https://www.cosmos.esa.int/gaia}),
processed by the Gaia Data Processing and Analysis Consortium (DPAC, 
\url{https://www.cosmos.esa.int/web/gaia/dpac/consortium}).
Funding for the DPAC has been provided by national institutions,
in particular the institutions participating in
the {\it Gaia} Multilateral Agreement. \\ \\
%\vspace{5mm}
%\newpage
%\vspace{5mm}
All the {\it HST} data sets used in this paper can be found in MAST.
Set 1 corresponds to And\,V ACS observations, set 2 to And\,V WFPC2 observations.
Set 3 corresponds to And\,VI ACS observations, set 4 to And\,VI WFPC2 observations.
\\ \\
%Set 1: \dataset[http://dx.doi.org/10.17909/r78a-kn48]{http://dx.doi.org/10.17909/r78a-kn48} \\
%Set 2: \dataset[http://dx.doi.org/10.17909/62eg-0f60]{http://dx.doi.org/10.17909/62eg-0f60} \\
Set 1: \dataset[https://archive.stsci.edu/doi/resolve/resolve.html?doi=10.17909/n5jp-ax84]{https://archive.stsci.edu/doi/resolve/resolve.html?doi=10.17909/n5jp-ax84} \\
Set 2: \dataset[https://archive.stsci.edu/doi/resolve/resolve.html?doi=10.17909/pveq-mt14]{https://archive.stsci.edu/doi/resolve/resolve.html?doi=10.17909/pveq-mt14} \\
Set 3: \dataset[https://archive.stsci.edu/doi/resolve/resolve.html?doi=10.17909/p6e0-dc79]{https://archive.stsci.edu/doi/resolve/resolve.html?doi=10.17909/p6e0-dc79} \\
Set 4: \dataset[https://archive.stsci.edu/doi/resolve/resolve.html?doi=10.17909/knfs-te70]{https://archive.stsci.edu/doi/resolve/resolve.html?doi=10.17909/knfs-te70} \\

\facilities{{\it HST}, MAST, {\it Gaia}}
\newpage
\bibliography{ms}{}

\end{document}